\documentclass[aps,prb, twocolumn,showpacs,preprintnumbers,superscriptaddress]{revtex4}

\usepackage{graphicx}% Include figure files
\usepackage{dcolumn}% Align table columns on decimal point
\usepackage{bm}% bold math

\begin{document}

\title{Effect of pressure on the magnetic, transport, and thermal-transport properties of the   electron-doped manganite CaMn$_{1-x}$Sb$_{x}$O$_{3}$}

\author{Yuh Murano}
\author{Michiaki Matsukawa} 
\email{matsukawa@iwate-u.ac.jp }
%\thanks{}
\author{Syuya Ohuchi} 
\author{Satoru Kobayashi} 
\affiliation{Department of Materials Science and Engineering, Iwate University , Morioka 020-8551 , Japan }
\author{Shigeki Nimori}
\affiliation{National Institute for Materials Science, Tsukuba 305-0047 ,Japan}
\author{Ramanathan Suryanarayanan}
\affiliation{Laboratoire de Physico-Chimie de L'Etat Solide,CNRS,UMR8182
 Universite Paris-Sud, 91405 Orsay,France}
\author{Keiichi Koyama}
\affiliation{Graduate School of Science and Engineering,Kagoshima University, Kagoshima 890-0065, Japan}
\author{Norio Kobayashi}
\affiliation{Institute for Materials Research, Tohoku University, Sendai  
980-8577, Japan}

\date{\today}

\begin{abstract}
We have demonstrated the effect of hydrostatic pressure on magnetic and transport properties, and thermal transport properties in   electron-doped manganites CaMn$_{1-x}$Sb$_{x}$O$_{3}$.
The substitution of Sb$^{5+}$ ion for Mn $^{4+}$site of  the parent matrix causes  one-electron doping with the chemical formula CaMn$^{4+}_{1-2x}$Mn$^{3+}_{x}$Sb$^{5+}_{x}$O$_{3}$ accompanied by a monotonous increase in unit cell volume as a function of $x$. Upon increasing the doping level of Sb, the magnitudes of both electrical resistivity and negative Seebeck coefficient are suppressed at high temperatures, indicating the electron doping. 
Anomalous diamagnetic behaviors  at $x=0.05$ and 0.08 are clearly observed  in field cooled dc magnetization.
The effect of hydrostatic pressure on dc magnetization is in contrast to the chemical pressure effect due to Sb doping.
%We suppose that the negative magnetization at low fields has some relationship with the appearance of a canted spin state, which is associated with the variation of $e_{g}$ orbital state through the local lattice distortion due to Sb doping. 
%We suppose that the tilting of MnO$_{6}$ octahedron due to Sb substitution causes a variation in the easy axis of magnetization and it establishes non collinear magnetic structure accompanied by the antisymmetric exchange interaction, resulting in the complicated magnetic properties including the negative magnetic moment. 
The dynamical effect of ac magnetic susceptibility measurement points to the formation of the magnetically frustrated clusters such as FM clusters embedded in canted AFM matrix. 
%In particular, we believe that the existence of the anomalous peak in the imaginary part of ac susceptibility under the zero field cooled condition is related to the  negative dc magnetization observed here.

%Non collinear magnetic structure is established by a variation in the easy axis of magnetization accompanied by the tilting of MnO$_{6}$ octahedron due to Sb substitution, and/or is stabilized by the antisymmetric exchange interaction, resulting in the complicated magnetic properties including the negative magnetic moment. 

%The application of external pressure on the x=0.05 sample changes the temperature dependent magnetization reversal at ambient pressure and results in a negative magnetization over a wide range of temperatures upon a zero-field cooling.
%The chemical pressure effect due to Sb doping suppresses the magnetic transition temperature, while
%the applied pressure enhances the magnitude of low field magnetization with its temperature.  
%These findings are also close to the variation of $e_{g}$ orbital state due to the local lattice distortion associated with the Sb doping.
\end{abstract}

%\pacs{75.47.Lx,75.50.Lk}
%\keywords{}
\renewcommand{\figurename}{Fig.}
\maketitle
\section{INTRODUCTION}
Manganese oxides with perovskite structure have been extensively investigated since the discovery of colossal magnetoresistance (CMR) effect.\cite{TO00}
The spontaneous insulator to metal transition and its associated CMR effect are well explained on the basis of the double exchange (DE) model between Mn$^{3+}$ and Mn$^{4+}$ ions.\cite{ZE51} Furthermore, the phase separation model, where the ferromagnetic (FM) metallic and antiferromagnetic (AFM) insulating clusters of competing electronic phases coexist, strongly supports experimental studies of manganites.\cite{DA01} The dynamic Jahn-Teller (JT) effect of Mn$^{3+}$ ions plays a crucial role in the physics of manganites.\cite{MI95}

CaMnO$_{3}$, the end member of Ca$_{1-x}$La$_{x}$MnO$_{3}$  system, undergoes a G-type antiferromagnetic transition around $T_N\sim 120$ K accompanied by a weak ferromagnetic component,\cite{MAC67} where each spin of Mn ions is antiparallel to its nearest neighbors of Mn. In recent years, the electron doped manganite system ($x<0.5$)\cite{CH96} has attracted much attentions because of the possibility of magnetoresistance effect. 
Neutron measurements on Ca$_{1-x}$La$_{x}$MnO$_{3}$ \cite{GRA03} have  revealed the formation of the nanometric-scale FM clusters isolated within a G-type AFM matrix for sufficiently low doping ($x<0.03$). Furthermore, for intermediate doping (0.03$\leq x\leq $0.14),  a canting of the G-type AFM structure occurs with the nanometric FM clusters. 
In addition to a large number of researches on the A-site substituted  electron doped manganites, 
it is shown that substituting of the Mn site of  CaMnO$_{3}$ with higher valence ions than 4+ such as
CaMn$^{4+}_{1-2x}$Mn$^{3+}_{x}$M$^{5+}_{x}$O$_{3}$, with M = Nb, Ta, V, Ru,\cite{RA00,PI03,SHA04,GU05,ANG06,ZHO09}and CaMn$^{4+}_{1-3x}$Mn$^{3+}_{2x}$M$^{6+}_{x}$O$_{3}$, with M = Mo, W,\cite{RA00,MAR01,PI03,MAI02,MIC07,ROZ08,MAR09} creates Mn$^{3+}$ ions, i.e., electrons leading to the CMR effect. 
The crystallographic and magnetic properties of CaMn$_{1-x}$Ru$_{x}$O$_{3}$ system ($x\geq 0.1$) allow us to apply the phase separated FM + AFM model to the ground state of this system.\cite{SHA04}
For CaMn$_{1-x}$Mo$_{x}$O$_{3}$ system with low doping of Mo ($x=0.04$), the low temperature magnetic ground state is better described by the canted AFM magnetic structure than by the phase separated state.\cite{ROZ08,MAR09} At higher doping levels of Mo, the charge ordered state is established within the parent matrix CaMnO$_{3}$.\cite{MAR01}   
It thus is interesting to examine the physical properties of the Mn-site substituted compositions for our understanding of electronic phase diagram of electron-doped manganites. 

In this paper, we demonstrate the effect of hydrostatic pressure on magnetic and transport properties, and thermal transport properties in  electron-doped manganites CaMn$_{1-x}$Sb$_{x}$O$_{3}$, in order to examine a relationship between lattice and spin. 
The physical pressure effect is a powerful probe to investigate the electronic states of manganese oxides varying the one-electron band width at doping level fixed because the application of external pressure gives rise to a shrinkage of Mn-O bond length and/or straightening of a Mn-O-Mn bond angle.\cite{KU97}  
There has been several studies on the effect of pressure on magnetism of calcium based electron doped manganites so far.\cite{MAR04} For CaMn$_{1-x}$Ru$_{x}$O$_{3}$ with $x=0.1$, the applied pressure dramatically suppresses the ferromagnetic phase accompanied by a rise of the magnetic transition temperature up to $\sim $14K. 

In previous works of slightly electron doped CaMnO$_{3}$ with B-site substitution, negative magnetization properties have been demonstrated.\cite{ANG06,MU10} 
The negative magnetization phenomena in manganites have been originally reported in compounds with two sublattices of Mn ions and rare-earth ions (Nd, Gd, Dy), such as  NdMnO$_{3}$\cite{BAR05,TRO06}, (La,Gd)MnO$_{3}$,\cite{HEM04} (Nd,Ca)MnO$_{3}$,\cite{TRO03} (Gd,Ca)MnO$_{3}$,\cite{PEN02} and (Dy,Ca)MnO$_{3}$.\cite{NO96,MO02}
Some of these studies were discussed on the basis of ferrimagnetic scenario leading to negative magnetization below a compensation temperature, where Mn and some rare-earths sublattices are antiferrmagnetically coupled.  Moreover, a phase separation model between the ferromagnetic clusters and the canted AFM matrix is proposed, in order to account for a possible origin of the negative magnetization in manganites.\cite{BAR05,TRO03}  
In addition to the B site substituted manganite, the negative magnetization was observed in some manganites without magnetic rare-earth ions, such as LaMnO$_{3}$ nano particles. \cite{MAR08}
Accordingly, the nature of the negative magnetization in manganites is one of crucial issues to be unveiled, which may be close to a phase segregated state.

Furthermore, we carry out the ac magnetic susceptibility measurements for CaMn$_{1-x}$Sb$_{x}$O$_{3}$, in order to examine the dynamic effect linked to magnetically frustrated properties. For spin glass or cluster glass system, a visible anomaly in the ac susceptibility appears upon lowering $T$ across freezing temperature of spins or clusters when the magnetic relaxation time becomes longer than the measuring time.\cite{MU81} 
In particular, we believe that a phase separated state realized in manganites is not consistent with a typical spin glass phase at low field.\cite{DE01} It thus is very intriguing to demonstrate a close relationship between static and dynamic responses of magnetization to the applied field, for our further understanding of  complicated magnetic behavior of the present samples. 
%For the present system studied here, magnetic ions are absent at the A site (Ca site) so that we exclude the ferrimagnet model consisting of two magnetic sublattices.
 
\section{EXPERIMENT}
\begin{figure}[ht]\includegraphics[width=10cm]{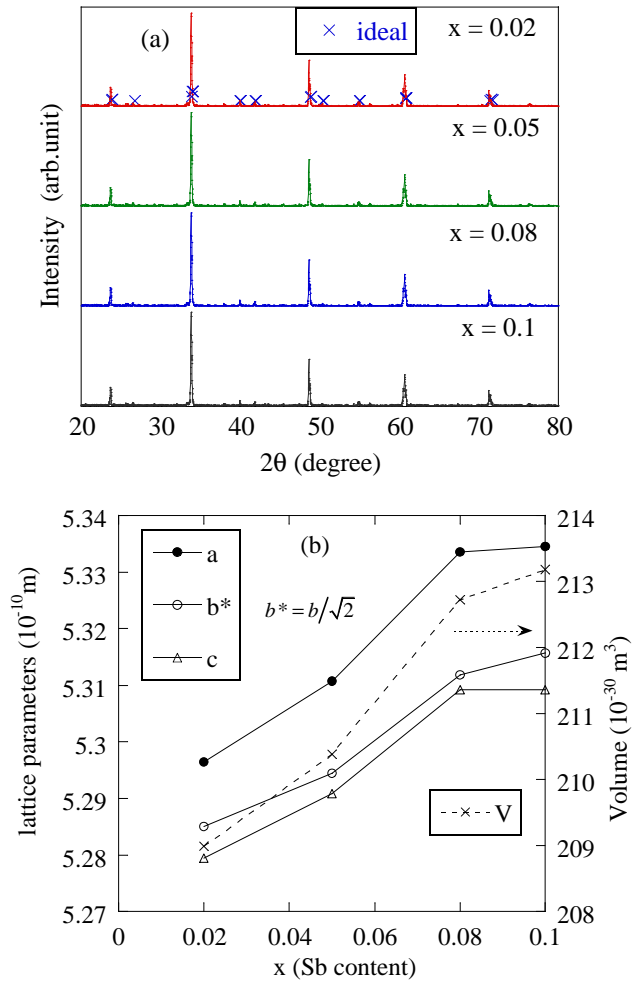}
\caption{(Color online) (a) X ray diffraction patterns of CaMn$_{1-x}$Sb$_{x}$O$_{3}$ ($x$=0.02, 0.05, 0.08 and 0.1) taken at room temperature. Cross marks denote intensity peaks of orthorhombic CaMnO$_{3}$. 
(b) The lattice parameters and unit cell volume as a function of Sb contents from 0.02 up to 0.1.
}\label{lattice}
\end{figure}%

Polycrystalline samples of CaMn$_{1-x}$Sb$_{x}$O$_{3}$ ($x$=0.02, 0.05, 0.08 and 0.1) were prepared with a solid-state reaction method. The   stoichiometric mixtures of  CaCO$_{3}$, Mn$_{3}$O$_{4}$, and Sb$_{2}$O$_{3}$ high purity powders were calcined in air at 1000 
$^{\circ}$C for 24 h. The products were then  ground and pressed into cylindrical pellets.  The pellets were finally sintered  at $1400\sim 1450$ $^{\circ}$C for 12 h. X-ray diffraction data revealed that all samples are almost single phase with orthorhombic structures ($Pnma$)(Fig.\ref{lattice}). The lattice parameters of the $x$=0.08 sample are $a=5.326$ \AA, $b=7.512$ \AA, and $c=5.310$ \AA, which is in fair agreement with a previous work.\cite{PO04} The cell parameters and unit cell volume increase with increasing the Sb doping because the ion radius of Sb$^{5+}$ (0.61\AA) is greater than the value of Mn$^{4+}$ (0.54\AA). In addition, the Mn$^{4+}$ ions are replaced by Mn$^{3+}$ ions  (0.645\AA) with one extra electron, which contributes to the increased cell volume. 
The electrical resistivity was measured with a four-probe method. Seebeck coefficient was determined from both measurements of a thermoelectric voltage and temperature difference along the longitudinal direction of the measured sample.  The thermal conductivity was collected with a conventional heat flow method.  The dc and ac magnetization measurement was carried out using commercial superconducting quantum interference device (SQUID) magnetometers both at Iwate Univ. and National Institute for Materials Science. The ac magnetic susceptibility measurement for $x=0.02$, 0.05, and 0.08 samples was measured  as a function of frequency and dc magnetic field at the ac magnetic field of 0.5 mT.  In particular,  to remove the influence of remanent magnetic field, the SQUID magnetometer with the option of magnet reset mode was used in low field measurements\cite{REM,MU10}.  
Hydrostatic pressures in magnetization and electrical resistivity measurements were applied by using a clamp-type CuBe cell up to 1 GPa. Fluorinert was used as a pressure transmitting medium. The magnitude of pressure was calibrated by the pressure dependence of the critical temperature of lead. Magnetoresistance effect was measured by using a superconducting magnet at the High Field Laboratory for Superconducting Materials, Institute for Materials Research, Tohoku University.

\section{Results and discussion}

\begin{figure}[ht]\includegraphics[width=10cm]{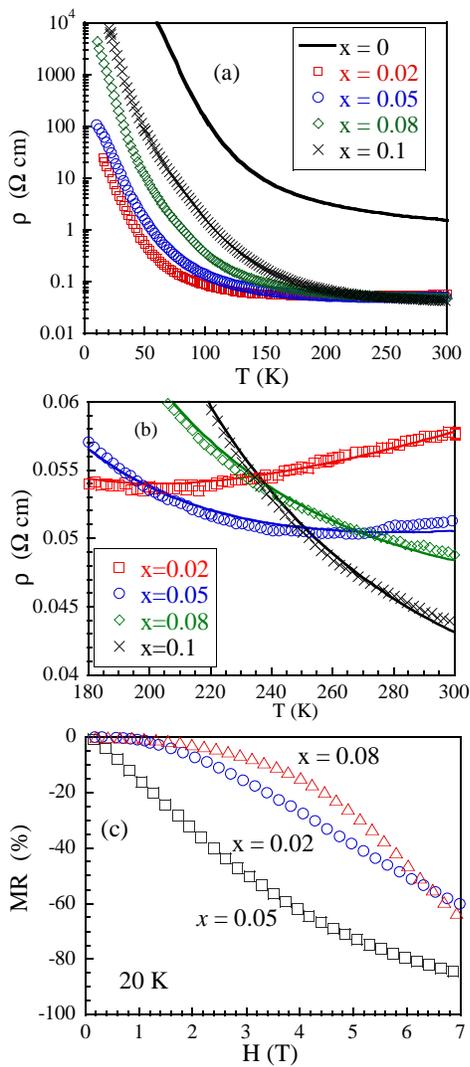}
\caption{(Color online) (a) Temperature variation of electrical resistivity $\rho$  for the CaMn$_{1-x}$Sb$_{x}$O$_{3}$ system ($x$=0.02, 0.05, 0.08, and 0.1). For comparison, the data of the parent compound are presented. (b) The  $\rho-T$ plot magnified between 180 K and room temperature. The solid lines denote the fits of the high-$T$ data by using a small polaron model. The activation energy $W$ is listed as a function of the Sb content in Table  \ref{lattice}.  (c) Magnetoresistance effect, $MR=[R(H)-R(0)]/R(0)\times100\%$ at 20 K  for the CaMn$_{1-x}$Sb$_{x}$O$_{3}$ system ($x$=0.02, 0.05, and 0.08). 
}\label{RT}
\end{figure}%

\begin{figure}[ht]\includegraphics[width=8cm]{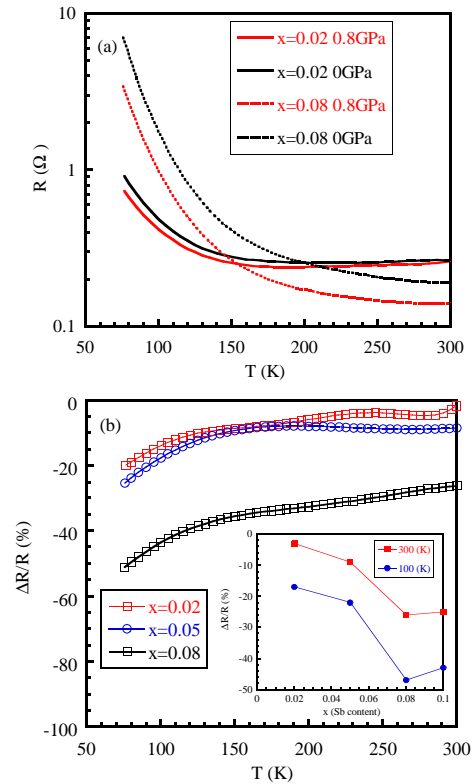}
\caption{(Color online) (a) Temperature variation of electrical resistance $R$  for the CaMn$_{1-x}$Sb$_{x}$O$_{3}$ system ($x$=0.02 and 0.08) at an applied pressure of 0.8 GPa. For comparison, the data at ambient pressure  are presented. (b)  $\Delta R/R =[R(T,0.8$GPa$)-R(T,0$GPa$)]/R(T,0$GPa) $\times100\%$  for the CaMn$_{1-x}$Sb$_{x}$O$_{3}$ system ($x$=0.02, 0.05, and 0.08). The inset displays $\Delta R/R$ as a function of Sb content at 100 K and 300 K.
}\label{RTP}
\end{figure}%

\begin{table}
\caption{\label{table1} Electrical transport characteristics of the CaMn$_{1-x}$Sb$_{x}$O$_{3}$ series. $\rho_{300K}$ represents the resistivity at 300 K. $W$ is calculated from the best fits of the high-$T$ transport data using the small polaron model. The  $MR$ under an applied field of 7 T and $\Delta R/R$  under an applied pressure of 0.8 GPa are estimated from the effect of magnetic field and pressure on the resistivity, respectively.(In detail, see the captions of Fig. \ref{RT} and Fig. \ref{RTP}). Magnetic transition temperature $T_m$ is estimated from the inflection point of the field cooled $M-T$ curve of Fig. \ref{MT} (a), (c), and (e).}
\begin{ruledtabular}
\begin{tabular}{ccccccccc}
Content&$\rho_{300K}$ &$W$& $MR_{20K}$&$\Delta R/R_{100K}$ &$T_m$&&&\\
$x$& ($\Omega $cm)& (meV)&($\%$)&($\%$)&(K)&&&\\
\hline
0&1.5&&&&120&&  \\
0.02&0.057&17.4&-60&-17&110&& \\
0.05&0.051&24.2&-85&-22&99&&   \\
0.08&0.049&33.9&-64&-47&88&& \\
0.1&0.044&45.2&&-43&38&&\\
\end{tabular}
\end{ruledtabular}
\end{table}

\subsection{Effect of magnetic field and external pressure on electrical transport }

The temperature variation of electrical resistivity $\rho$  for the CaMn$_{1-x}$Sb$_{x}$O$_{3}$ system  is shown in Fig. \ref{RT} as a function of the Sb content. The Sb substitution for Mn site up to $x=0.1$ gives rise to a substantial decrease in high temperature $\rho$(300 K) from 1.5 $\Omega$cm at the parent $x=0$ sample through 0.057 $\Omega$cm at the lightly doped $x=0.02$ sample down to 0.044 $\Omega$cm at the intermediately doped $x=0.1$ sample, indicating the carrier doping into Mn-site. 
 All samples exhibit semiconducting behaviors upon decreasing temperatures except for a metallic conduction at high temperatures for the Sb light doping.  The $\rho-T$ curve of the $x=0.02$ and 0.05 samples remains a metallic like character in the paramagnetic region down to $\sim$ 200 K and 250 K, respectively. 
On the other hand, the magnitude of the resistivity of the $x$=0.08 and 0.1 is enhanced at lower temperatures since the further doping of Sb ion breaks some of conduction paths along the Mn-O-Mn network, resulting in reinforcing carrier localization. In previous works,\cite{JA96,MAI98} it is reported that the high temperature transport in doped manganites is described by a small polaron hopping model.  Now, we try to fit the resistivity data at high temperatures by using such an expression as $\rho=AT$exp($W/kT$), where the fitting parameter $W$ represents the activation energy.  As listed in Table \ref{table1}, with increasing the Sb concentration, the value of the activation energy shows a gradual increase from 17.4 meV at $x=0.02$ up to 45.2 meV at $x=0.1$ in spite of a monotonous suppression in the resistivity vs Sb content at high temperatures. In the preceding section of Thermal transport, we will discuss the transport mechanism in the CaMn$_{1-x}$Sb$_{x}$O$_{3}$ system.
We notice that the resistivity data of parent CaMnO$_{3}$ observed here are, both in temperature dependence and magnitude, similar to the  $\rho-T$ curve of the stoichiometric composition without oxygen defects, CaMnO$_{3-\delta }$ ($ \delta=0 $).\cite{ZE99}  

Next, let us examine the negative magnetoresistance effect on  the electron doped manganite samples with 
$x=0.02$, 0.05, and 0.08 as shown in Fig. \ref{RT}(c). At the $x$=0.05 sample, the gigantic magnetoresistance attaining  $-85\%$ at 20 K under the applied field of  7 T is observed. For the other samples, we obtain the MR ratio of $\sim -60\%$. 
The MR data are comparable to those reported in a previous work\cite{RA00} on CaMn$_{1-x}$M$_{x}$O$_{3}$ (M=Nb,Ta) system. Theses findings strongly indicate that the spin polarized ferromagnetic metal clusters are established by the applied field and the field-induced delocalized state is realized within the samples studied. 

Finally, let us show in Fig. \ref{RTP} the effect of pressure on the electrical resistance $R$ as a function of temperature between 77 K and 300 K  for the CaMn$_{1-x}$Sb$_{x}$O$_{3}$ ($x$=0.02 and 0.08). For comparison, the data at ambient pressure  are presented. 
The applied pressure of 0.8 GPa on the $x=0.08$ sample suppresses the magnitude of $\rho$ from 25$\%$ at 300 K down to about 50 $\%$ at lower temperatures around 80 K upon decreasing temperature.
The effect of pressure on the electrical transport observed here seems to be more enhanced below near the magnetic transition temperatures $T_{m}$=110 K at $x=0.02$ and 88 K at $x=0.08$, where  $T_{m}$ is determined  from the inflection point of the magnetization data of Fig. \ref{MT}. 
The pressure dependence of the resistance $\Delta R/R$ (the inset of Fig. \ref{RTP}) exhibits a maximum at the heavily doped sample of  $x=0.08$, which is consistent with the effect of strong pressure on the magnitude of magnetization below $T_{m}$.(see Figs. \ref{MT}(e) and \ref{MT}(f)) 

\subsection{Thermal transport (Seebeck coefficient and thermal conductivity) }
\begin{figure}[ht]\includegraphics[width=9cm]{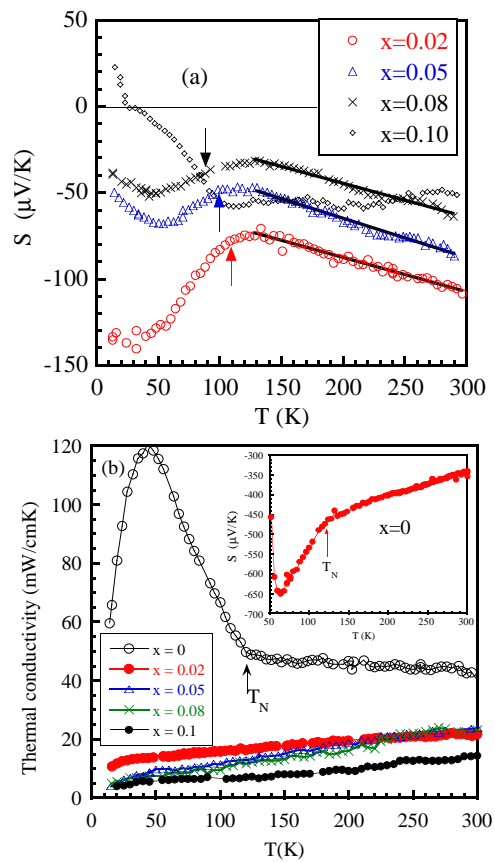}
\caption{(Color online) (a) Temperature variation of  Seebeck coefficient $S$ for the CaMn$_{1-x}$Sb$_{x}$O$_{3}$ system ($x$=0.02, 0.05, 0.08 and 0.1). The solid lines denote the $T$ linear fits to the $S(T)$ data in the temperature region above 130 K.
The arrows denote the magnetic transition temperatures determined from the magnetization data.
(b) Thermal conductivity $ \kappa $ of CaMn$_{1-x}$Sb$_{x}$O$_{3}$ ($x$=0, 0.02, 0.05, 0.08, and 0.1). For comparison, the $\kappa $ data of the parent compound CaMnO$_{3}$ are presented. The inset also represents the  $S$ data of parent CaMnO$_{3}$. 
%Seebeck coefficient $S$ vs  $T^{1/2}$ for the $x$=0.02,0.05 and 0.08 samples.  The solid lines correspond to fits by the 3D VRH model.  The inset represents the  $S$ data of parent CaMnO$_{3}$. 
%In the case of x=0.05, we have $A=27.2$ $\mu$V/K and $B=6.54$ $\mu$V/$K^{3/2}$ for $p=1/2$
%($A=-19.2$ $\mu$V/K and $B=0.23$ $\mu$V/$K^{2}$ for $p=1$),where $S(T)=A-BT^{p}$. 
%Thermal conductivity $ \kappa $ of CaMn$_{1-x}$Sb$_{x}$O$_{3}$ ($x$=0, 0.02, 0.05 and 0.1). For comparison, the $\kappa $ data of the parent compound CaMnO$_{3}$ are presented. 
}\label{SK}
\end{figure}%

Seebeck coefficient $S$ is suppressed upon increasing Sb doping as shown in Fig. \ref{SK}(a). For all samples studied, $S$ shows a negative value over a wide range of temperatures, strongly suggesting electron doping into the parent matrix. 
In the case of the substitution of Sb$^{5+}$ ion for Mn$^{4+}$ site, the Mn$^{4+}$ ions are replaced by Mn$^{3+}$ ions with one extra electron.
For lower doped samples ($x$=0.02 and 0.05), the local maximum observed in $S$ is located near the magnetic transition temperature as mentioned above. With further doping of Sb, the local maximum suppressed in $S$ of $x$=0.08 is still located around 130 K, but it has little relationship with the weakly magnetic transition near 90 K.  
We notice a common feature in the $T$-linear dependence of Seebeck coefficient for three samples ($x$=0.02, 0.05 and 0.08) between 130 K and room temperature, although their $T$ dependences of the resistivity are quite different over the same temperature range as shown in Fig. \ref{RT}(b). In our previous section, we demonstrated that the high-$T$ resistivity is well fitted by using a small polaron model. The corresponding expression for Seebeck coefficient is given in the form of 
\[S(T)=\frac{k_{B}}{e}\left\{\frac{E_{S}}{k_{B}T}\right\}+S_{\infty },\]
where $E_{S}$ and $S_{\infty }$ represent the thermal activation energy and the spin entropy in the high temperature limit, respectively.\cite{JA96}  
In Fig. \ref{SK}, we note that the $1/T$ law is violated at high temperatures. 
The electrical resistivity of the present samples is greatly influenced by grain boundaries in comparison to Seebeck measurement, giving rise to little reliable information about the electronic states.\cite{SE} 
In fact, such disagreements between these transports of polycrystalline oxide materials have been pointed out as far.\cite{FI00,MAE87}
 
In previous works\cite{MAI98,FI00} of electron doped manganites, the $T$-linear dependence of $S(T)$ has been discussed on the basis of the conventional metal model, or Culter and Mott (CM) model.
The former is given in  the form of 
%\[S(T)=\pi^{2} k_{B}/3e\{k_{B}TN(E)/n+const.\}_{E=E_{F}},\]
\[S(T)=\pi^{2} \frac{k_{B}}{3e}\left\{k_{B}T\frac{N(E)}{n}+const.\right\}_{E=E_{F}},\]
where $N(E)$ is the density of states and $n$ is the carrier density.\cite{MAI98}
The latter formula for the CM model is expressed as 
\[S(T)=-\pi^{2} \frac{k_{B}}{3e} \left\{ k_{B}T\frac{\partial \ln (\mu_{0} N(E))}{\partial E} -\frac{\partial W}{\partial E} \right\} _{E=E_{F}},\]
%\[ \left\{ \frac{\partial W}{\partial E} \right\} \] 
where $\mu =\mu_{0} $exp$(-W(E)/k_{B}T )$ is an activated mobility and $ W(E)$ is an activation energy.\cite{CU69} The CM model for the random hopping system  well describes a random distribution of localized states of electrons around Fermi level.
The typical fitting parameters $N(E)/n$ and $\{\partial $ln$(\mu_{0} N)/\partial E \}^{-1}$, for the conventional and CM models are listed in Table \ref{TableS}.  The fitted result shows that $N(E)/n$ ratio is almost independent of the nominal composition. %from the $S(T)$ data observed and   the ratio of the number of total Mn ions to that of Mn$^{3+}$ ion $n(Mn)/n(Mn^{3+})$ from 
The magnitudes of $\{\partial $ln$(\mu_{0} N)/\partial E \}^{-1}$ 
are similar to those of the A-site substituted manganite Ca$_{1-x}$La$_{x}$MnO$_{3}$ ($x$=0.017 and 0.033).\cite{FI00}  For the insulating Li$_{1+x}$Ti$_{2-x}$O$_{4}$ sintered samples, it has been reported that $S(T)$ is proportional to temperature and the $S(T)$ behavior is analyzed using the CM model.\cite{MAE87} 
The Sb substitution induced lattice disorder influences the electronic states of the parent matrix, causing the validity of the CM model.  Concerning the B-site substituted manganites,  there is a common trend in the temperature dependence of $S(T)$ between CaMn$_{1-x}$W$_{x}$O$_{3}$($x=0.02$ and 0.04) and the present samples with $x=0.02$, 0.05, and 0.08.\cite{MIC07} 
\begin{table}
\caption{\label{TableS} Thermal transport characteristics of the CaMn$_{1-x}$Sb$_{x}$O$_{3}$ series. $S_{300K}$ represents Seebeck coefficient at 300 K. $N(E)/n$ is calculated from the best fits of the high-$T$ data using the conventional model. $n(Mn)/n(Mn^{3+})$ is the ratio of the number of total Mn ions to that of Mn$^{3+}$ ion.
We obtain from the inverse slopes of $S$,$\{\partial $ln$(\mu_{0} N)/\partial E \}^{-1}$, by using the CM model.}
\begin{ruledtabular}
\begin{tabular}{ccccccccc}
Content&$S_{300K}$ &$\frac{N(E)}{n}$&$\frac{n(Mn)}{n(Mn^{3+})}$&
 $\left( \frac{\partial \ln (\mu_{0} N(E))}{\partial E} \right)^{-1}$&&&&\\
$x$& ($\mu $V/K)& (1/eV)&&(eV)&&&&\\
\hline
0&-340&&&&&&  \\
0.02&-107&8.2&50&0.12&&& \\
0.05&-84&9.3&20&0.11&&&   \\
0.08&-63&8&12.5&0.13&&& \\
0.1&-51&&&&&&\\
\end{tabular}
\end{ruledtabular}
\end{table} 

%In addition, we try to examine the thermal transport mechanism of our samples, using by 
%the Mott's variable-range-hopping (VRH) model\cite{SHK84}.  For the 3D VRH case, the corresponding form is reduced  to the form as $S(T)\propto T^{1/2} $ when Mott's activation energy is in the same order of $T$ , \cite{ZVY91}.
%Between a local maximum temperature around 120 K and room temperature, we also obtain a better  fit of the experimental data to the 3D VRH conduction than to the thermally activated $T$ dependence (Fig.\ref{S}(b)).
%Within our fitting procedures, there is no obvious differences between the $T$ linear and $T^{1/2}$ dependence of $S(T)$.   
%However, the corresponding electrical resistivity is not described by $\rho\propto $ exp$(T_{0}/T)^{1/4}$ for 3D VRH model, where $T_{0}$ is Mott's activation energy. 

%These findings indicate that the electrical resistivity of the present samples is greatly influenced by grain boundaries and reveals no intrinsic properties of the electron doped manganite system as reported in a previous work\cite{FI00}.

For $x$=0.1, the $S(T)$ data remain a nearly constant from room temperature down to near 100 K, then increase rapidly and finally exhibit a positive value across the horizontal axis at lower temperatures.
Such an unusual behavior in  $S(T)$ has been reported both in A-site substituted system Ca$_{1-x}$Sm$_{x}$MnO$_{3}$($x$=0.2)\cite{HE99} and B-site system CaMn$_{1-x}$W$_{x}$O$_{3}$($x\geq 0.07$)\cite{MIC07}, indicating a holelike character of charge carriers.

%\begin{figure}[ht]\includegraphics[width=8cm]{FigKT.ps}
%\caption{Thermal conductivity $ \kappa $ of CaMn$_{1-x}$Sb$_{x}$O$_{3}$ ($x$=0, 0.02, 0.05 and 0.1). For comparison, the $\kappa $ data of the parent compound CaMnO$_{3}$ are presented. 
%}\label{K}
%\end{figure}%
Finally, let us show in Fig. \ref{SK} the temperature variation of  thermal conductivity in CaMn$_{1-x}$Sb$_{x}$O$_{3}$ ($x$=0, 0.02, 0.05, 0.08, and 0.1).
The thermal conduction for all samples studied here is carried by acoustic phonons because the electron contribution is negligible by an estimation from the electric resistivity data using the Wiedemann-Franz law.  
Upon decreasing temperature crossing the antiferromagnetic temperature T$_{N}$,  a sharp increase in  $ \kappa $  of parent CaMnO$_{3}$ is observed as reported in previous works.\cite{HE99,CO02}  We believe that 
the thermal conductivity in the paramagnetic phase  is strongly suppressed by phonon scattering due to nanoscale strains generated by spin correlations.\cite{CO02} This finding is thus  explained   by a rapid reduction of the phonon scattering when  the AFM long range order is established below T$_{N}$.
The  light doping of  Sb$^{5+}$  strongly suppresses the magnitude of thermal conductivity from 120 mW/cmK (50K) at the pure x=0 sample down to 14 mW/cmK at $x$=0.02 (Fig.\ref{SK}). The Sb$^{5+}$  doping removes the Mn$^{4+}$ ion and instead produces the Mn$^{3+}$ ion for Mn sites which is the Jahn-Teller active ion with one e$_{g}$-electron. It is believed that the local lattice distortion due to JT effect causes phonon scattering, which is close to the depressed thermal conduction.\cite{MA03} In addition, the lattice deformation  due to the Sb doping with its larger ion radius affects the neighboring Mn$^{3+}$O$_{6}$ octahedron, resulting in some variation of the orbital-state of e$_{g}$-electron through the local JT effect.

\subsection{Effect of pressure and magnetic field on dc magnetization}
\begin{figure}[ht]\includegraphics[width=9cm]{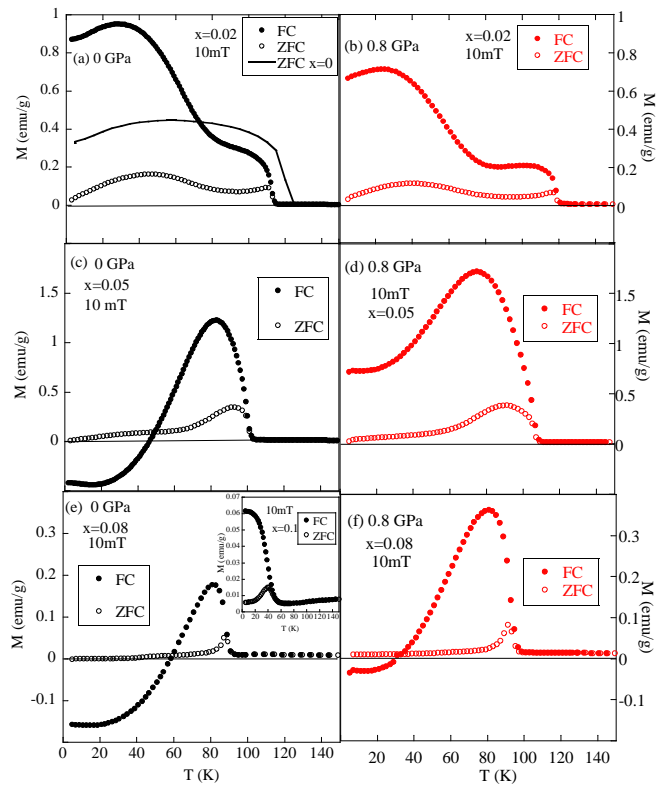}
\caption{(Color online) Temperature dependence of zero field cooled and field cooled  magnetization of the Sb substituted CaMn$_{1-x}$Sb$_{x}$O$_{3}$ recorded at the magnetic field of 10 mT. In the case of   ambient pressure, (a) $x=0.02$, (c) $x=0.05$, and (e) $x=0.08$. In the case of 0.8 GPa, (b) $x=0.02$, (d) $x=0.05$, and (f) $x=0.08$. 
For comparison, the ZFC data of the parent sample are given in (a). In the inset of (e), the $M-T$ curve with the $x=0.1$ sample is presented at 0 GPa. 
}\label{MT}
\end{figure}%
\begin{figure}[ht]\includegraphics[width=9cm]{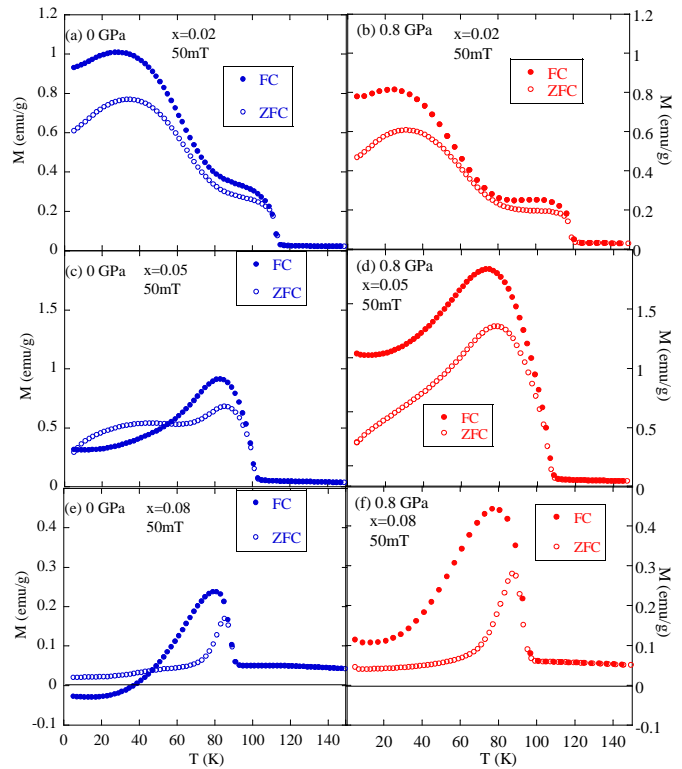}
\caption{(Color online) ZFC and FC magnetization curves of  the Sb substituted CaMn$_{1-x}$Sb$_{x}$O$_{3}$ ($x=0.02$, 0.05, and 0.08) collected at 50 mT. For ambient pressure, we get (a) $x=0.02$, (c) $x=0.05$, and (e) $x=0.08$. For applied pressure of 0.8 GPa, (b) $x=0.02$, (d) $x=0.05$, and (f) $x=0.08$. 
}\label{MT50}
\end{figure}%

%\begin{figure}[ht]\includegraphics[width=10cm]{FigMTP.ps}
%\caption{Effect of pressure on zero field cooled and field cooled  magnetization of  the Sb substituted CaMn$_{1-x}$Sb$_{x}$O$_{3}$ under 0.8 GPa. (a) x=0.02, (b) x=0.05, (c) x=0.08. 
%}\label{MTP}
%\end{figure}%

Let us show in Figs. \ref{MT}(a), \ref{MT}(c), and \ref{MT}(e) the zero-field cooled (ZFC) and field cooled (FC) magnetization data of the Sb substituted CaMn$_{1-x}$Sb$_{x}$O$_{3}$ measured under an applied magnetic field of 10 mT at 0 GPa. First of all,  temperature dependent magnetization reversal appears in the FC curves of both $x=0.05$ and 0.08, as it has been reported in the V doped CaMn$_{1-x}$V$_{x}$O$_{3}$ manganites.\cite{ANG06} 
However, lowering temperature down to 2 K, the ZFC data of both samples show no negative value. 
Upon increasing the Sb content up to $x=0.1$, a large difference between ZFC and FC curves is observed but such negative magnetization phenomenon vanishes.   
%The ZFC curve seems to be a mirror image of the FC curve with respect to the horizontal axis. 
%For $x=0.05$, the ZFC curve starts from its positive value, then shows a monotonic decrease upon increasing $T$,and finally reaches a negative minimum around 80 K.  The negative magnetization is limited between the intermediate and magnetic transition temperatures.We notice similar characters in the magnetization data of both the x=0.05 and 0.08 samples.
%We note that the ZFC data for all samples show no negative magnetization.
%a large divergence between ZFC and FC curves appears at low and intermediate temperatures, but such negative magnetization phenomenon vanishes even in FC modes.

%suggesting spin-glass like magnetic state. 
The magnetic transition temperature is suppressed from 110 K at $x=0.02$, through 99 K at  $x=0.05$, down to 38 K at$x=0.1$ due to the Sb doping because the Sb$^{5+}$ ion is non magnetic one with the closed shell of $4d^{10}$.
The substitution of non magnetic Sb ions for Mn sites introduces electron doping, forming ferromagnetic clusters, but destroys magnetic ordering between Mn ions within the parent matrix.

The FC data of $x=0.08$ indicates that  the antiparallel component of  the canted AFM spins to the applied field is  stabilized by the Sb substitution. 
The spontaneously canted magnetic moment contributing to the negative magnetization is close to a variation of the local easy axis which is caused by  the local lattice distortion of Sb substitution in comparison to the Sb free sample.  If the number of canted spin clusters contributing to the antiparallel component is dominant over that of these clusters contributing to the parallel one along the applied field, we then obtain the negative magnetization. 
Furthermore, we need to point out the significant roles of the Dzyaloshinsky-Moriya  interaction which prefers canted spin arrangements. If the antisymmetric exchange  interaction between nearest neighbor Mn ions is dominant, such  mechanism then stabilizes non collinear spin configuration, leading to complicated magnetic behaviors.
In order to account for the negative magnetization observed in the lightly doped Nd$_{1-x}$Ca$_{x}$MnO$_{3}$ series, the low-temperature spin reorientation is explained in the framework of a two-phase model, where the samples consist of exchange coupled ferromagnetic and weak ferromagnetic phases.\cite{TRO03} The reorientation of magnetic moment of Nd ions in both phases plays a crucial role on the magnetization reversal through the f-d exchange interaction between the Nd and manganese sublattices. It seems that the proposal model is ruled out because the magnetic rare earth ion is absent in the present system. 
%The temperature induced magnetization reversals in the orbital ordered YVO$_{3}$  have been reported.\cite{REN98}
%It is true that our data  are, in temperature dependence, quite similar to those phenomena. However, the CaMnO$_{3}$ system with low doping yields no orbital ordering of e$_{g}$ electrons and it is hard to directly apply the possible mechanism for the appearance of negative magnetization in YVO$_{3}$ system. 
%Nevertheless, the DM interaction and local lattice transformation associated with JT effect probably give a clue to reveal the complex magnetic properties.

%For YVO$_{3}$ , the temperature induced magnetization reversals in low magnetic fields are interpreted as two coupled spin canting mechanisms, where one mechanism is concerning an antisymmetric Dzyaloshinski-Moriya exchange interaction and another is single ion magnetic anisotropy\cite{REN98}. 
%is explained on the basis of the effect of spin canting of AFM domains on the magnetization\cite{ZE99}.  The anisotropy energy, retaining spins along the AFM easy axis, is lower than the energy gained by reorientation of AFM spins from antiparallel direction to canting arrangement 
%We note that the parent composition CaMnO$_{3}$  is a G-type AFM insulating state with a weak ferromagnetic moment, which is probably caused by a minor defect of oxygen or from Dzyaloshinsky-Moriya interaction preferring the spin-canted state\cite{NE00,CH07}.

Next, we attempt to measure the effect of pressure on the magnetization for CaMn$_{1-x}$Sb$_{x}$O$_{3}$, to examine a relation between lattice distortion and spin arrangement (Figs. \ref{MT}(b), \ref{MT}(d), and \ref{MT}(f)). 
The application of hydrostatic pressure up to 0.8 GPa enhances $T_{m}$ by 8$\sim $9 K  at $x=0.02$ and 0.05, and a stable increase in $T_{m}$ reaches about 5 K at the $x=0.08$ sample.  
We expect that a shrinkage of lattice parameters makes stronger  a super-exchange interaction between nearest neighbors $t_{2g}$ spins of  Mn$^{4+}$ ions, resulting in stabilization of the G-type AFM magnetic structure.
The magnitude of magnetization is not largely changed at $x=0.02$ and 0.05, but the FC magnetization of the $x=0.08$ sample exhibits a rapid rise in $M_{max}$  from 0.18 emu/g at 0 GPa up to 0.36 emu/g at 0.8 GPa, which is responsible for a sharp increase of FM clusters induced by pressure with a stable rise of $T_{m}$. 
Both the ZFC and FC curves at $x=0.02$ under 0.8 GPa are similar to those under ambient pressure
but the magnitude of magnetization is suppressed by the application of pressure. 
%The value of  $T_{m}$ increases with pressure as  $dT_{m}/dP\sim6$ K/GPa for $x=0.02$ which is in good agreement with the pressure coefficient ($dT_{m}/dP=4.8\pm 0.5$ K/GPa) of the parent CaMnO$_{3}$.\cite{MAR04} 
The application of external pressure on the $x=0.05$ sample changes the temperature induced magnetization reversal observed at 0 GPa and results in positive FC curve, approaching the ambient magnetization curves of the low doped sample ($x=0.02$).  At further Sb content of  $x=0.08$, the negative magnetization of the FC curve at 0.8 GPa are limited at lower temperatures below $\sim $30 K.
However, it seems that the ZFC curves of all samples remain qualitatively unchanged even under the applied pressure. 
The orthorhombic distortion due to Sb doping is suppressed by application of pressure, resulting in a similar magnetic behavior to the lower doped case under ambient pressure.
%
%In a similar way, the data of $x=0.08$ sample under pressure approach the temperature variation of magnetization of $x=0.05$ without pressure. These findings are also common character in the case of  the effect of pressure at 50 mT,  as shown in Fig. \ref{MT50}, where the ZFC and FC magnetization data of CaMn$_{1-x}$Sb$_{x}$O$_{3}$ at applied field of 50 mT under 0 GPa and 0.8 GPa are given.
At the applied field of 50 mT, the negative magnetization disappears except for the low temperature region of $x=0.08$ at ambient pressure as shown in Fig. \ref{MT50}. 

Here, we make remarks about a notable difference between the magnetic properties of the $x=0.1$ sample and other ones
($x=0.02$, 0.05, and 0.08). 
As mentioned above, the magnetic transition temperature $T_m$ is monotonously decreased from 110 K at $x=0.02$ through 99 K at $x=0.05$ to 88 K at $x=0.08$ upon further Sb doping.  At a maximum content of $x=0.1$, $T_m$ is strongly depressed down to 40 K. These findings indicate that the magnetic interaction working between manganese ions  is considerably suppressed beyond $x=0.08$.  
For the  $x=0.1$ sample, the nominal content of $e_g$ electron is taken as a maximum in the all samples studied.  In fact, the room temperature resistivity and corresponding Seebeck coefficient of the $x=0.1$ sample exhibit the lowest values.  However, upon lowering temperatures, the suppression of magnetic coupling working among the Mn ions around each Sb ion probably becomes dominant over the carrier doping effect and causes a strongly reduced magnetization accompanied by a remarkable decrease of  $T_m$.  In addition to it, we point out the giant pressure effect on the magnitude of magnetization of the $x=0.08$ sample in applied fields of 10 mT and 50 mT as depicted in Fig. \ref{MT}(d) and Fig. \ref{MT50}(d).  In particular, the weaken magnetic interaction due to Sb doping under ambient pressure is strengthened under applied pressure of 0.8 GPa, giving the remarkable increase in the magnetization by a factor of about 2.  The critical content is located near $x=0.08$, separating negative and normal magnetic tendencies.

\begin{table}
\caption{\label{table3} Effect of pressure on magnetic transition temperature $T_m$ of the CaMn$_{1-x}$Sb$_{x}$O$_{3}$ system ($x$=0.02, 0.05, and 0.08 ). $T_m$ and $T_m^{p}$ are determined from the inflection of $M-T$ curves of Fig.\ref{MT} under 0 GPa and 0.8 GPa, respectively. The $M_{max}$ and $M_{max}^{p}$ denote  a maximum in FC magnetization curves below the magnetic transition temperatures under 0 GPa and 0.8 GPa, respectively.  
 $\Theta $ and $\mu_{exp}$ represent the Curie-Weiss temperature and the effective magnetic moment  per one manganese ion estimated from a high temperature linear fit of $1/M-T$ under ambient pressure. 
For comparison, the effective magnetic moment $\mu_{cal}$is given. (In detail,see the text) }
%We have the value of $M$ at 5 T from the low-$T$ isothermal magnetization  in Fig.\ref{MH}(c). }
\begin{ruledtabular}
\begin{tabular}{ccccccccc}
&$T_m$&$T_m^{p}$& $M_{max}$&$M_{max}^{p}$&$\Theta $&$\mu_{exp}$&$\mu_{cal}$&\\
$x$& (K)& (K)&(emu/g)&(emu/g)&(K)&($\mu_{B}$/Mn)&($\mu_{B}$/Mn)&\\
\hline
0&120&&&&&&3.87  \\
0.02&110&119&0.95&0.71&-119&3.51&3.85 \\
0.05&99&107&1.23&1.70&-1&3.41&3.83   \\
0.08&88&93&0.18&0.36&28&3.70&3.81 \\
0.1&38&&&&&&3.80\\
\end{tabular}
\end{ruledtabular}
\end{table} 
\begin{figure}[ht]\includegraphics[width=8cm]{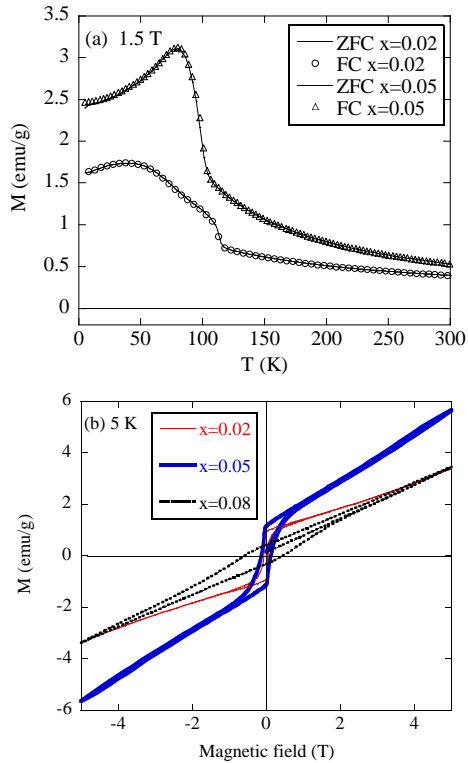}
\caption{(Color online) (a) Temperature dependence of ZFC and FC magnetization curves of the CaMn$_{1-x}$Sb$_{x}$O$_{3}$ ($x=0.02$ and $0.05$)  samples under an applied field of 1.5 T. (b) Magnetization curves  $M$ vs $H$ of $x=0.02$, 0.05, and 0.08  at 5 K.
}\label{MH}
\end{figure}%

Let us determine from the $1/M-T$ data taken at 0.1 T (not shown here), the Curie-Weiss temperature and the effective magnetic moment, for CaMn$_{1-x}$Sb$_{x}$O$_{3}$ system ($x$=0.02, 0.05, and 0.08 )  as listed in Table \ref{table3}.    
The Curie-Weiss temperature $\Theta $ is estimated from a high temperature linear fit of $1/M-T$ by using the relation $ \chi=C/(T-\Theta)$. The Curie constant $C$ gives the effective magnetic moment $\mu_{exp}$ per one manganese ion. For comparison, the effective magnetic moment $\mu_{cal}$ is calculated as a function of nominal Sb content 
from $\mu_{eff}^2=(1-2x)\mu^2$(Mn$^{4+}$)+$x\mu^2$(Mn$^{3+}$).  For free ions in the high spin configuration, we have, $\mu $(Mn$^{4+}$) and $\mu $(Mn$^{3+}$) , 3.87$\mu_{B}$ and 4.90$\mu_{B}$, respectively. Here, we assume that $\mu =2\sqrt{S(S+1)}$ where the spin quantum number $S$ = 3/2 for Mn$^{4+}$ and $S$ = 2 for Mn$^{3+}$. 
In the paramagnetic regime, we obtain $\Theta $ from the negative value of -119 K at $x$=0.02  to the positive ones of 28 K at $x$=0.08, which is indicative of a qualitative variation in the magnetic interaction from antiferromagnetic to ferromagnetic character. Furthermore, the   effective magnetic moment shows a slight increase from 3.51 $\mu_{B}$ at $x$=0.02 up to 3.70 $\mu_{B}$ at $x$=0.08. This finding seems to point to, the existence of ferromagnetic fluctuation or the formation of FM small clusters, in the paramagnetic phase of the Sb doped sample above the magnetic transition temperature. 
However, the resultant magnetic parameters are not always consistent with the low temperature magnetization suppressed at $x$=0.08.  The suppression of magnetic interaction due to Sb doping as discussed above and its related magnetic frustration probably prevent establishment of a large FM cluster or a long range FM order at low temperatures.
   
In a relatively high field of 1.5 T, the temperature variation of magnetization for $x=0.02$ and 0.05 shows no magnetic hysteresis between ZFC and ZC curves as displayed  in Fig. \ref{MH}(a).
%Figure \ref{MH}(b) shows the $M$($H$) loop of the $x=0.02$ sample measured at 5 K. The initial $M$($H$) curve of the $x=0.02$ sample recorded at 5 K is changed from its negative to positive sign when the applied field exceeds 0.1 T.   
The $M$($H$) data with a small hysteresis do not saturate even at 5 T  and rises linearly with increasing $H$ as shown in Fig. \ref{MH}(b), indicating the AFM matrix with  small FM phase. 
We understand from both  Figs. \ref{RT}(c) and \ref{MH}(b) that the larger magnetization at high fields  corresponds the stronger magnetoresistance effect. 
A linear extrapolation of $M$($H$) curve at 5 K to $H=0$ gives a spontaneous magnetization $M_{s}$ from 0.97 emu/g at $x=0.02$ through 1.2 emu/g at $x=0.05$  down 0.42 emu/g at $x=0.08$, revealing the existence of small residual magnetization for all samples studied. 

Finally, we suppose the valence fluctuation of Sb ion, to account for discrepancies in the effective moment between the experimental and calculated values as listed in Table \ref{table3}. In a recent study of CaMn$_{1-x}$Ru$_{x}$O$_{3}$\cite{ZHO09}, X-ray absorption measurements reveal the presence of the mixed valence states of Ru ion.  In the intermediate samples (0.1$\leq x\leq $0.9), it has been reported that the valence states of Ru$^{5+}$ and Ru$^{4+}$ coexist with the mixed valence of Mn ion. 
However, X-ray photoelectron spectroscopy analysis of La$_{0.9}$Sb$_{0.1}$MnO$_{3}$\cite{DU04} indicates that the valence state of Sb ion is +5.  Thus, we believe that Sb$^{5+}$ exists predominantly in the low doping region of Sb.

\subsection{Frequency and dc magnetic field dependences of ac magnetic susceptibility}

\begin{figure}[ht]\includegraphics[width=8cm]{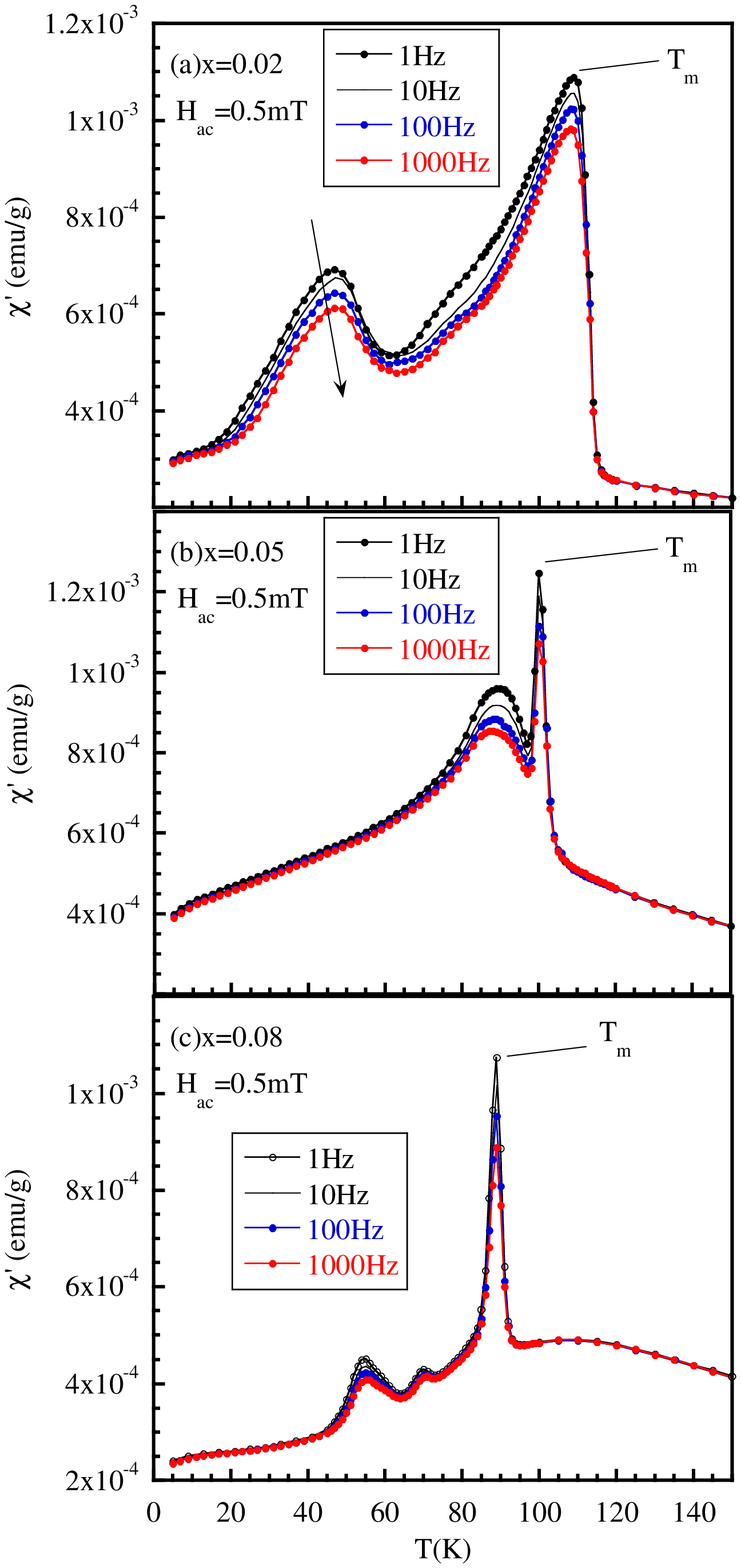}
\caption{(Color online) The real part $\chi'$  of ac magnetic susceptibility of CaMn$_{1-x}$Sb$_{x}$O$_{3}$ collected at zero dc magnetic field with frequency ranging 1 Hz up to 1 kHz. 
(a) $x=0.02$, (b) $x=0.05$, and (c)$x=0.08$. The amplitude of the ac magnetic field $H_{ac}$ was set to be 0.5 mT. The arrows point to the direction of increasing frequencies.  
}\label{ACRe}
\end{figure}%
\begin{figure}[ht]\includegraphics[width=8cm]{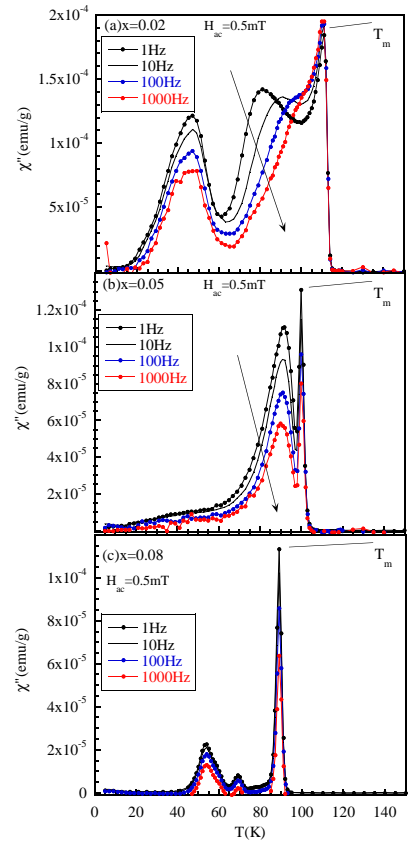}
\caption{(Color online) The imaginary part $\chi''$  of ac magnetic susceptibility of CaMn$_{1-x}$Sb$_{x}$O$_{3}$  as a function of frequency from 1 Hz to 1 kHz at $H_{dc}$ = 0 T.($H_{ac}$ = 0.5 mT)  (a) $x=0.02$, (b) $x=0.05$, and (c)$x=0.08$. 
}\label{ACIm}
\end{figure}%

\begin{figure}[ht]\includegraphics[width=9cm]{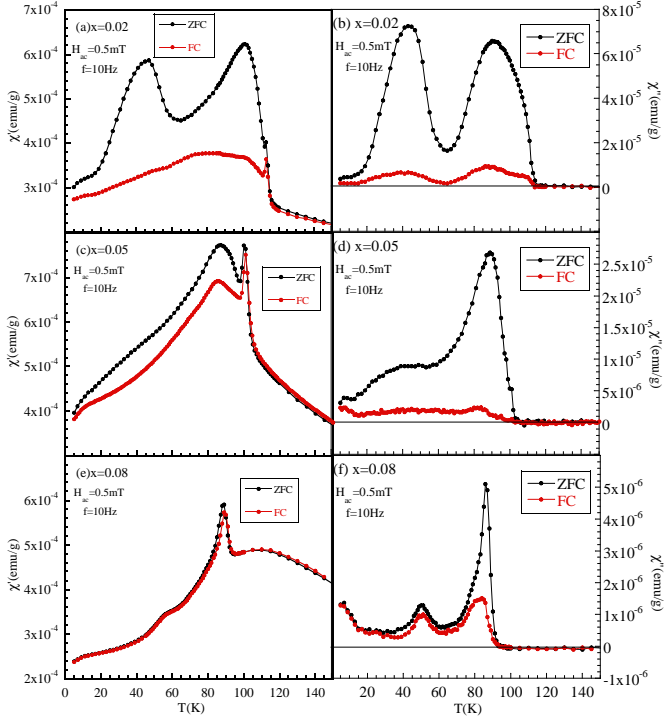}
\caption{(Color online) The real and imaginary parts of ac magnetic susceptibility  of 
CaMn$_{1-x}$Sb$_{x}$O$_{3}$ measured at 10 Hz under a superimposed dc field ($H_{dc}$=10 mT). For $x=0.02$, (a) $\chi'$ and (b)$\chi''$. For $x=0.05$, (c) $\chi'$ and (d)$\chi''$. For $x=0.08$, (e) $\chi'$  and (f)$\chi''$. 
 The ac magnetization data are recorded as a function of temperature under ZFC and FC conditions. 
}\label{ACFC}
\end{figure}%
\begin{figure}[ht]\includegraphics[width=8cm]{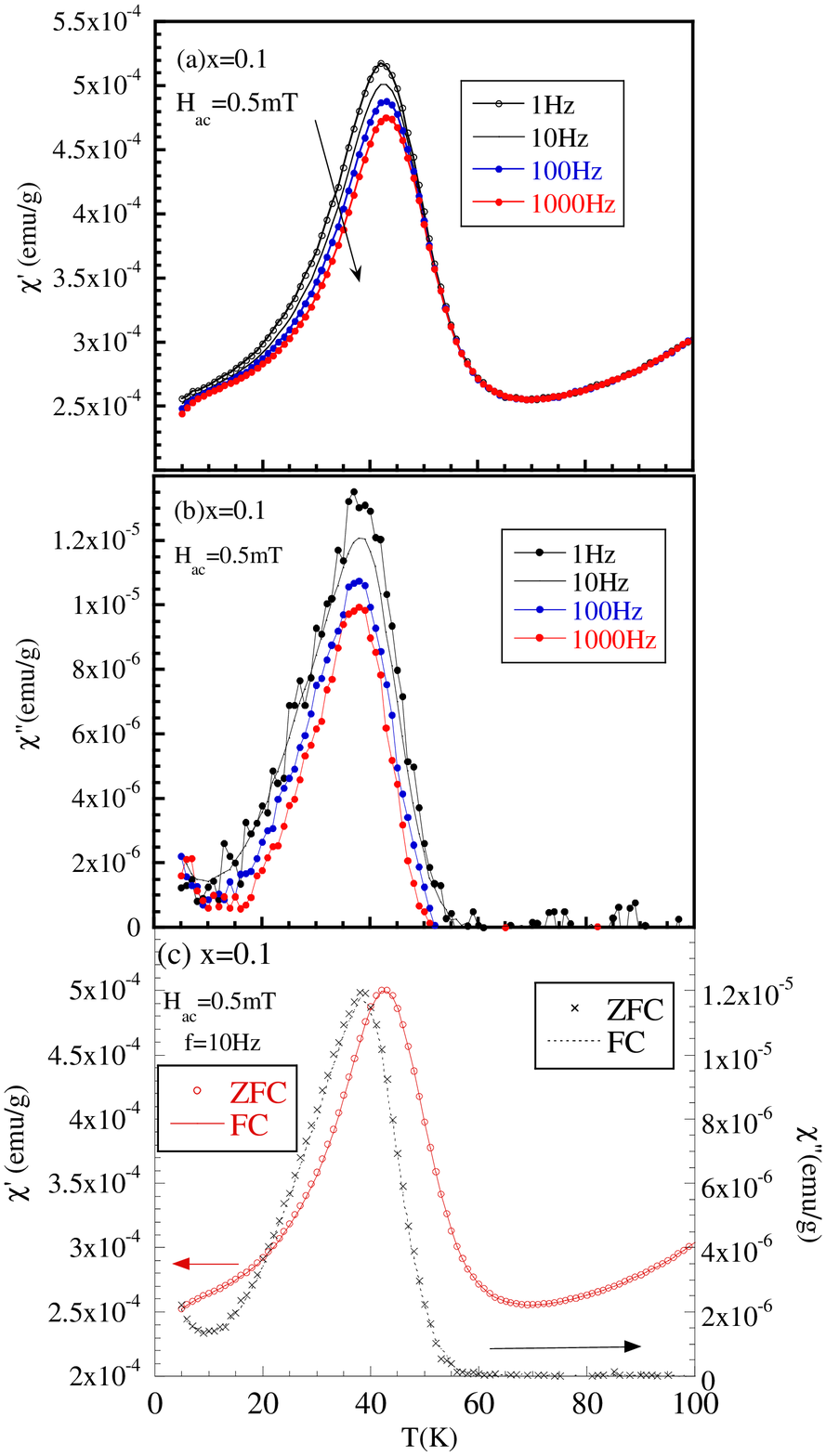}
\caption{(Color online)   The real and imaginary parts of ac magnetic susceptibility of CaMn$_{0.9}$Sb$_{0.1}$O$_{3}$ a function of frequency from 1 Hz to 1 kHz at $H_{dc}$ = 0 T.($H_{ac}$ = 0.5 mT). For $x=0.1$,(a) $\chi'$ and (b)$\chi''$. For comparison, the ac magnetization data recorded as a function of temperature under ZFC and FC conditions are given in (c) ($H_{dc}$=10 mT).
}\label{AC0.1}
\end{figure}%

Furthermore, we carry out the ac magnetic susceptibility measurements for $x=0.02$, 0.05, and 0.08 samples, in order to examine the dynamic effect linked to magnetically frustrated properties (Figs. \ref{ACRe} and \ref{ACIm}). 
The real and imaginary parts of the ac susceptibility, $\chi'$ and $\chi''$,  are registered at zero dc magnetic field with frequency $f$ ranging 1 Hz up to 1 kHz. For $x=0.02$, a sharp transition is noticed around $T_{m}=110$ K, which well agrees with  the dc magnetic measurements. A maximum peak of $\chi'$ curves shows no shift in temperature  with increasing $f$ but a second peak at lower temperature exhibits a small shift towards high temperatures, indicating the presence of a magnetic frustration.  The magnitude of $\chi'$(T) curves is suppressed with frequency over a wide range of temperatures. In addition to the two peaks observed in $\chi'$,  a third peak in  $\chi''$  appears at low frequencies, its peak then shifts towards higher temperatures with increasing $f$, and finally it is merged in a maximum peak at high frequencies. The new peak around 80 K at 1 Hz  strongly depends on frequency. The imaginary component representing magnetic energy dissipation is close to the low field dc magnetic hysteresis  between ZFC and FC.
For $x=0.05$, the maximum peak  is located at  $T_{m}=100$ K, while the second one observed  at higher temperature of 90 K is accompanied by a slight frequency shift towards low temperatures.  It seems that the frequency shift of the second peak 
in $\chi'$ and $\chi''$ curves for both $x=0.02$ and 0.05 is indicative of the signature of  a spin-glass like magnetic property. However, a substantial decrease of peak in $\chi''$ with increasing frequency is qualitatively in contrast to the behavior of conventional spin glass system,\cite{MU81} as previously reported in phase separated Pr$_{0.7}$Ca$_{0.3}$MnO$_{3}$.\cite{DE01}  
The dynamical effect of $\chi'$ has some relationship with the formation of the magnetically frustrated clusters such as FM clusters embedded in canted AFM matrix.  
%However, we have no idea whether these findings give a direct evidence for the negative magnetization. 
On the other hand, in the case of $x=0.08$, the second peak in $\chi'$ and $\chi''$ curves is located near 55K and it shows a weak frequency dependence as shown in Figs.\ref{ACRe} and \ref{ACIm}.  
We notice that the second peaks of $x=0.02$ and 0.05 are located around 47 K and 90 K, respectively, corresponding to the maximum of the ZFC dc magnetization.
However, the second peak of $x=0.08$ has no corresponding maximum in the ZFC dc curve (Fig. \ref{MT}(c)).  
We note that the ac susceptibility of the parent CaMnO$_{3}$ exhibits no magnetic peak except for its maximum peak associated with long range  magnetic ordering.\cite{MAI98} Moreover, for the B site substituted CaMn$_{1-x}$Mo$_{x}$O$_{3}$($x=0.04$), the imaginary part $\chi''$ shows only a rapid peak at 105 K around  the magnetic transition temperature and no signal at low temperatures.\cite{MAR09}  For the low doped CaMn$_{1-x}$Ru$_{x}$O$_{3}$($x=0.06$),
both $\chi'$ and $\chi''$ exhibit neither frequency dependence nor enhanced peak below $T_{m}$.\cite{SHA04} 
This finding predicts that the low temperature ground state of the light Ru substituted CaMnO$_{3}$ is better explained by the complex canted magnetic structure than by the phase separated model. 
For A site substituted Ca$_{0.9}$La$_{0.1}$MnO$_{3}$, a much larger hysteresis in the dc magnetization between ZFC and FC curves, and the frequency effect on the ac susceptibility demonstrate that its low temperature state is considered as a cluster glass one without long range ferromagnetism.\cite{MAI98}  

Now, under zero field cooled  and field cooled conditions, we display in Fig. \ref{ACFC} the ac susceptibility measurements of $x=0.02$, 0.05, and 0.08 with a superimposed dc field ($H_{dc}$=10 mT). 
First of all, the steep peaks in both $\chi'$ and $\chi''$ around $T_{m}$ in the absence of dc field are strongly suppressed  under the application of dc field but the second peaks of $x=0.02$ and 0.05 still remain stable.  
%In general, the characteristic temperature of the ac magnetization peak is considered as the freezing temperature of spins or magnetic clusters. 
The collapse of a maximum peak at zero dc field is caused by the application of low dc field since the applied low field strongly suppresses the magnetic fluctuation associated with the magnetic transition.\cite{MUK96} 
Next, a magnetic divergence in the ac magnetization  between ZFC and FC curves is visible at $x=0.02$,  but upon increasing the Sb content, both ZFC and FC ac curves at $x=0.08$ become reversible and exhibit no clear differences 
except for the high temperature region near $T_{m}$.
Finally, for the $x=0.08$ sample the maximum peak in FC $\chi''$ of Fig. \ref{ACFC}(f) is observed around 85 K, at the temperature where the FC dc curve reaches a maximum as shown in  Fig. \ref{MT}(e).
If we assume that the maximum peak in FC $\chi''$  is related to the formation of magnetically frustrated clusters  associated with a phase separated state,  we can not identify the second peak with corresponding signature  in dc magnetization curve.  
However,  the anomalous peak is located near the characteristic temperature pointing to the dc magnetization reversal.
%The ac  $\chi''$  data  at  $x=0.05$ show no clear anomaly around the temperature crossing zero but a local maximum is detected near $ \sim$ 50 K. 
Accordingly, we suppose that the existence of the anomalous peak in $\chi''$ of $x=0.08$ has some relationship with the negative magnetization phenomena observed here.
For comparison, the ac susceptibility data of the $x=0.1$ sample are presented in Fig. \ref{AC0.1}. We notice frequency dependence of a maximum peak in both $\chi'$ and $\chi''$ around 40 K corresponding to the ZFC dc peak in the inset of Fig.\ref{MT}(e), indicating the strong evidence for magnetically frustrated state.  

\section{SUMMARY}
 We have demonstrated the effect of hydrostatic pressure on magnetic and transport properties, in   electron-doped manganites CaMn$_{1-x}$Sb$_{x}$O$_{3}$. In addition, thermal transport properties (Seebeck coefficient and thermal conductivity) of the  CaMn$_{1-x}$Sb$_{x}$O$_{3}$ system  have been examined as a function of $T$. Furthermore, the ac magnetic susceptibility measurements for $x=0.02$, 0.05, and 0.08 samples have been performed, in order to examine a close relation between the dynamic effect linked to magnetically frustrated properties  and the static dc  magnetization. 

The substitution of Sb$^{5+}$ ion for Mn $^{4+}$site of  the parent matrix causes  one-electron doping with the chemical formula CaMn$^{4+}_{1-2x}$Mn$^{3+}_{x}$Sb$^{5+}_{x}$O$_{3}$ accompanied by a monotonous increase in unit cell volume as a function of $x$. 

Upon increasing the doping level of Sb, the magnitudes of both electrical resistivity and Seebeck coefficient are suppressed at high temperatures, indicating the electron doping. The CM model applied to the random hopping system gives a better fit to Seebeck coefficient at higher temperatures. The  light doping of  Sb$^{5+}$  strongly suppresses the high thermal conductivity of the parent sample through the local lattice distortions.   

The anomalously diamagnetic behaviors  at $x=0.05$ and 0.08 are clearly observed  in the field cooled spontaneous magnetization.
The magnetization curves under the applied pressure of 0.8 GPa at $x=0.05$ and 0.08 exhibit similar behaviors to those of the $x=0.02$ and 0.05 samples without pressure, respectively.
We expect that these findings are close to some change of the local easy axis of magnetization due to the local lattice distortion induced by the Sb doping, in comparison to the case of the A site substitution.
A notable difference in the magnetization curves between the $x=0.1$ sample and other ones is attributed to the weaken magnetic interaction working among the Mn ions around Sb ion. 
%Non collinear magnetic structure is established by a variation in the easy axis of magnetization accompanied by the tilting of MnO$_{6}$ octahedron due to Sb substitution, and/or is stabilized by the antisymmetric exchange interaction, resulting in the complicated magnetic properties including the negative magnetic moment. 

The dynamical effect of ac magnetic susceptibility measurement has some relationship with formation of the magnetically frustrated clusters such as FM clusters embedded in canted AFM matrix. 
%A significant divergence in the anomalous peaks of $\chi''$  between the ZFC and FC curves of the $x=0.02$ sample indicates that the negative dc magnetization under the ZFC condition is metastable in comparison to the magnetic state under the FC condition or in the presence of relatively high dc magnetic fields.
In particular, we suppose that the existence of the anomalous peak in FC $\chi''$ of $x=0.08$ is related to the negative magnetization observed here.
% since the ac magnetization curves of the B site substituted CaMn$_{1-x}$Mo$_{x}$O$_{3}$($x=0.04$) and CaMn$_{1-x}$Ru$_{x}$O$_{3}$($x=0.06$) without negative dc magnetization behavior show no clear anomaly  below the magnetic transition temperature.

%It is desirable to carry out the neutron diffraction measurement on polycrystalline Sb-doped CaMnO$_{3}$, for our further understanding of anomalously magnetic properties.
%We have to take into account for the tilting of the easy axis, the antisymmetrical exchange interaction,  and the magnetic frustration, for our understanding of the complicated magnetic properties. 

\begin{acknowledgments}

 This work was partially supported by a Grant-in-Aid for Scientific Research from Japan Society of the Promotion
of Science. 
\end{acknowledgments}


\begin{thebibliography}{30}
\bibitem{REM}In our previous work\cite{MU10}, we presented the low field magnetization measurements by using the SQUID magnetometer.  Recently, we noticed that the remanent magnetic field in the superconducting magnet affects the temperature variation of the magnetization in the zero field cooled runs.  In fact, we checked it by using the SQUID magnetometer with the function of magnet reset mode, to reduce the remanent field below about 2G.  Separately, we tried to perform the magnetization measurements with the SQUID by using the fluxgate device for further reduction of the remanent value below 1 mG.  The former result was in good agreement with the latter. As a result, we expect that the apparent $''$ZFC$''$ scan in the previous measurements is a field cooled one under the negative remanent field in the case of switching off the magnet reset.
\bibitem{SE}Seebeck coefficient  $S$ is a sensitive probe to examine the electronic state of disordered crystals or highly inhomogeneous materials. $S$ is less affected by grain boundaries of polycrystalline samples than the electrical conductivity measurement since the thermal current applied to the sample is hard to be disturbed by the grain boundaries in comparison to the electric current.
\end{thebibliography}
\end{document}